\documentclass[12pt]{article}
\usepackage{amssymb,amsmath}
\usepackage[square,
comma,
sort&compress]{natbib}
\include{epsf}

\usepackage{graphicx}
\usepackage{amsfonts}

\newcommand{\e}{{\rm e}}
\newcommand{\ii}{{\rm i}}
\newcommand{\dd}{{\rm d}}
\newcommand{\eqn}[1]{(\ref{#1})}

\def\appendix#1{\addtocounter{section}{1}\setcounter{equation}{0}
\renewcommand{\thesection}{\Alph{section}}
\section*{
\thesection\protect\indent \parbox[t]{11.715cm} {#1}}
\addcontentsline{toc}{section}{Appendix\thesection\ \ \ #1} }
\newcommand{\complex}{{\mathbb C}} 
\newcommand{\zed}{{\mathbb Z}} 
\newcommand{\real}{{\mathbb R}} 

\def\vev#1{\left\langle #1\right\rangle}
\newcommand{\tr}[1]{\:{\rm tr}\,#1}

\newcommand{\be}{\begin{equation}}
\newcommand{\ee}{\end{equation}}
\newcommand{\beq}{\begin{equation}}
\newcommand{\eeq}{\end{equation}}
\newcommand{\bea}{\begin{eqnarray}}
\newcommand{\eea}{\end{eqnarray}}
\def\beqa{\begin{eqnarray}}
\def\eeqa{\end{eqnarray}}

\newcommand{\del}{\partial}

\newcommand{\eq}{\begin{equation}}
\newcommand{\eqa}{\begin{eqnarray}}
\newcommand{\en}{\end{equation}}
\newcommand{\ena}{\end{eqnarray}}

\begin{document}
\begin{titlepage}
\begin{flushright}

\baselineskip=12pt
DSF--26--2008\\
ICCUB-08-134
\hfill{ }\\
October 2008
\end{flushright}

\begin{center}

\baselineskip=24pt

{\Large\bf The Structure of Spacetime and Noncommutative Geometry}

\baselineskip=14pt

\vspace{1cm}

{\bf Fedele Lizzi\footnote[0]{Talk given at the workshop: \emph{Geometry, Topology, QFT and Cosmology},
Paris, 28-30 May 2008. To appear in the proceedings.}}
\\[6mm]
{\it Dipartimento di Scienze Fisiche, Universit\`{a}
di Napoli {\sl Federico II}}\\ and {\it INFN, Sezione di Napoli}\\
{\it Monte S.~Angelo, Via Cintia, 80126 Napoli, Italy}
\\[6mm]
{\it Institut de C\'incies del Cosmos,
Universitat de Barcelona,
Mart\'i Franq\'us 1, 08193 Barcelona, Catalonia, Spain
}\\{\small\tt
fedele.lizzi@na.infn.it}
\\[10mm]

\end{center}

\vskip 2 cm

\begin{abstract}
We give a general and nontechnical review of some aspects of noncommutative geometry as a tool to understand the structure of spacetime. We discuss the motivations for the constructions of a  noncommutative geometry, and the passage from commutative to noncommutative spaces. We then give a brief description of Connes approach to the standard model, of the noncommutative geometry of strings and of field theory on noncommutative spaces. We also discuss the role of symmetries and some possible consequences for cosmology.
\end{abstract}

\end{titlepage}

\section{Introduction}
In this contribution I will give a general, and personal, overview
of some attempts that physicists and mathematicians are making to
understand the structure of spacetime at extremely small
distances. The tool used for the description of spacetime at the
Planck length scale is what is called \emph{Noncommutative
Geometry}. This is a name which covers a wide range of slightly
different approaches, unified by the theme that physics at
small scales requires a modifications of the usual geometrical
concepts. As I said this review is personal, and I refer to the
literature for more a systematic and complete treatment of the
subject. Standard texts on noncommutative geometry are~\cite{connes, landi, ticos}.

I will start with a definition of geometry, for which I cannot
find an authority higher than \textsl{Wikipedia} \cite{Wikipedia}:

\noindent {\bf Geometry} {\sf (Greek
$\gamma\varepsilon\omega\mu\varepsilon\tau\rho\iota\alpha$; geo =
earth, metria = measure)} {\em is a part of mathematics concerned
with questions of size, shape, and relative position of figures
and with properties of space.}

Geometry is at the hearth of several physical theories, including
classical mechanics, special and general relativity. Classical mechanics, in its simplest form of point
particle mechanics, can be seen as the geometrical theory of phase
space, or of the position-velocity space, in the Hamiltonian or
Lagrangian framework. A (pure) state is a point in this space and
time evolution is given by a Hamiltonian vector field,
generated by an Hamiltonian function and a Poisson bracket. In
special relativity the configuration space is generalized to the
Minkowski space-time, which becomes the set of all point-like
\emph{events}. This construction is then further generalized to
general relativity, the theory of gravitation. In this case the
space becomes curved, but still the event is a point-like object.

In these theories the geometry used is the mostly usual one, based
on points, lines, tangent vectors etc. The change imposed by
quantum mechanics, even remaining in the one particle framework,
is dramatic. The phase space undergoes a drastic transformation
imposed by Heisenberg uncertainty principle:
\be
\Delta x \Delta p \geq \frac{\hbar}2
\ee
That there must be uncertainty can be heuristically seen with the
so-called Heisenberg's microscope. In its barest simplicity it can
be described in the following simple way\footnote{For a more 
quantitative description see for example the book
\cite{EspositoMarmoSudarshan}. This book is also useful in general
for understanding the relations between quantum and classical
mechanics.}. If one wants to measure the position of a microscopic
particle the only way to proceed is to send a ray of light on it,
the ray will be scattered by the particle, and collected by a
microscope. Light rays have a particular wavelength $\lambda$, and
it is impossible to measure the position of the particle below the
the wavelength of the photon. Hence to measure with better and
better precision the position of the particle, it is necessary to
use photons with shorter and shorter wavelength. But the joint
presence of the two fundamental constants $\hbar$ (Planck's
constant) and $c$ (speed of light) means that the photons have
energy $E=\hbar /\lambda$ and momentum $p=\hbar c /\lambda$,
therefore ``small'' photons are also very energetic. To ``see''
the particle we have to scatter a photon against the particle, and
then collect it in the microscope. But a very energetic photon
will have given a portion of its momentum to the photon, and this
quantity is unknown, thus rendering uncertain the measure of its
momentum. The concept of point of the phase space therefore loses
its meaning.

Naturally this is only a caricature of the uncertainty relation. To fully understand this issue  a new framework is built: quantum
mechanics. Position $x$ and momentum $p$ become self-adjoint
\emph{noncommuting} operators on a Hilbert space, and the
uncertainty principle is a consequence of the relation
$[x,p]=i\hbar\delta^i_j$. Noncommutativity is intimately
related with the loss of the concept of point. The quantum phase
space is not a geometric object in the classical sense, it is a
space not made of points, still it has a geometrical structure
which should become the usual one in the limit when $\hbar\to 0$.
Geometry, even the classical one, remains a fundamental tool of
investigation also for quantum mechanics,
but clearly the central role played by the point structure of the
space has to be relinquished, or generalized.

\section{The need for a different kind of geometry \label{se:need}}
\setcounter{equation}{0}

In ordinary quantum mechanics configuration space is still
described by the ordinary geometry. Points in space
still make sense. If we are willing to pay the price of losing
information about momentum we can still measure position with
arbitrary precision. Is it legitimate to expect the usual geometry
to hold to all scales in configuration space? When gravity enters
the game the answer must be no. There are several arguments to
support this no. One of them~\cite{DoplicherFredenhagenRoberts, Doplicher} is a paraphrases of the
Heisenberg microscope. Very roughly speaking (we refer to the
literature for details), if one wants to measure distances of the
order of Planck's length, then it is necessary to concentrate a
large amount of energy in a small volume. But when the amount of
energy is too large  a black hole is formed, effectively screening
the short distance. There are more argument coming from different
theoretical  frameworks which lead to the same qualitative
conclusion, see for example the review~\cite{Garay}. In
general the localization of events becomes at least
problematic, if not impossible, at the fundamental length obtained
with the fundamental constants of nature, including gravitational
constant: Planck's length
$\ell_P=\sqrt{\frac{G\hbar}{c^3}}=1.6~10^{-33}~cm.$

On the other side it is known that the construction of a quantum
field theory of gravitation (a quantum gravity) is hampered by the
ultraviolet problem of the nonrenormalizable divergences due to
the graviton. This again is a small distance problem, and it is
natural to think that the two issues are related. The root of the problem is that
we still lack the theory of quantum gravity! There are two major
candidates to quantum gravity (not the
only ones). One is string theory~\cite{Polchinski}, the
other is loop quantum gravity~\cite{Rovelli}. In both cases
spacetime receives novel interpretations. For strings the
coordinates which describe it are fields of a two-dimensional
conformal field theory. In loop quantum gravity the change is
more radical, spacetime is substituted by a network of spins. Both
theories have their supporters and detractors, roughly reflecting
the fact that both can claim successes but are still affected by
problems. Noncommutative geometry is intimately connected with
both these approaches, and the overlaps are considerable. We will
comment on some of them (for the string case) in the
following.

\section{From commutative to noncommutative geometries}
\setcounter{equation}{0}

To understand noncommutative geometry and the tools it uses it is
necessary to first make a step backward, and consider
\emph{commutative geometry} in a way which makes the
generalization natural. A good starting point is a series of
theorem by Gel'fand and Naimark~\cite{FellDoran, ticos} which
establish a one to one correspondence between \emph{Hausdorff
topological spaces} on one side, and \emph{commutative
$C^*$-algebras} on the other. A Hausdorff topological space is a
space in which points can be separated, i.e.\ given two points, it
is always possible to find two disjoint open sets each of them
containing one of the points, the space is separable.

\newcommand{\norm}[1]{\|#1\|}

A $C^*$-algebra $\mathcal{A}$ is an associative algebra over the
complex numbers $\complex$, which has a structure of Banach
algebra with a norm $\norm{a}$ and an involution (generalized
complex conjugation) ${}^*$ which satisfies the properties {\small
\begin{enumerate}
\item $\norm{a} \geq 0~, ~~~ \norm{a} = 0 ~\iff ~ a = 0~$
\item $\norm{\alpha a} = |\alpha|\,\norm{a}$
\item $\norm{a+b} \leq \norm{a} + \norm{b}$
\item $\norm{ab} \leq \norm{a}\, \norm{b}$
\item $(a^{*})^* = a$
\item $(ab)^* = b^* a^*$
\item $(\alpha a + \beta b)^* = \bar{\alpha} a^* + \bar{\beta}b^*$
\item $\norm{a^*}=\norm{a}$
\item $\norm{a^*a}=\norm{a}^2$ \label{cstarnorm}
\item It is
complete with respect to the norm.
\end{enumerate}
}

To each Hausdorff topological space is possible to associate a
commutative $C^*$-algebra in a canonical way; the algebra of
continuous complex valued function. The remarkable result is that
the converse is also true, i.e.\ every commutative $C^*$-algebra
is the algebra of continuous complex valued functions on some
Hausdorff topological space. The proof is constructive: the points are given by the one-dimensional
representations of the algebra. The topology can be reconstructed by considering
limit of a sequence of representation $x_n$ to
be the representation $x=\lim_nx_n$ with the property that
\be
\lim_n x_n(a)=x(a)\ \forall a\in\mathcal{A}
\ee
The set of one-dimensional representations coincides with the set
of \emph{pure states}. A state $\phi$ is a positive
normalized\footnote{Positive means that $\phi(a^*a)\geq 0$. The
norm is defined as the supremum of $\phi(a^*a)$ for $\|a\|=1$.}
functional from the algebra into the complex numbers. A pure state
is a state which cannot be expressed as the convex sum of two
other states.

The generalization can be done by ``simply'' considering
\emph{noncommutative} $C^*$-algebras. In this case however there
is no more a duality. In some cases the algebra can be still
associated to an underlying topological space, for example
the algebra of matrix valued functions on a manifold. But in
some other cases the underlying set of separated points simply
does not exist. This is the case of the algebra of position and
momentum of ordinary quantum mechanics\footnote{Technically,
since position and momentum are unbounded operators, to generate a
$C^*$-algebra one has to consider their exponentiated version of unitary operators.}. This is the original noncommutative
geometry, and in the presence of Heisenberg uncertainty clearly
shows that the concept of point of a quantum phase space is not
useful. We will come back to this kind of noncommutative geometry
later.

For several years here has been an effort, spearheaded by Alain
Connes, to write some sort of ``dictionary" to translate all
concepts of ordinary geometry in an algebraic form, without
reference to the underlying point structure, so to enable a
generalization to the noncommutative case. There are some key
ingredients of this effort and we mention just a few
of them without details, just to give the flavour of this activity.

An important result is again due to Gel'fand and Naimark, together
with Segal. They have shown that every $C^*$-algebra, commutative
or not, can be represented by an algebra of bounded operators on a
suitable Hilbert space. The proof is again constructive. Roughly
speaking the Hilbert space is built from the algebra itself and
particular (cyclic) state $\rho$. The latter provides the
possibility to define an inner product
\be
(a,b)=\rho(a^*b)
\ee
The Hilbert space is then the quotient of the algebra itself by
the ideal of states for which $\rho(a^*a)=0$, completed in the
Hilbert space norm. Note that the Hilbert space norm is different
form the $C^*$ norm. We refer again to the literature for details.

In order to define the extra geometrical ingredients we need a
generalized Dirac operator $D$. This operator enables the
translation of the metric and differential structure of spaces in
an algebraic form. The $D$ operator is a not necessarily bounded,
self adjoint operator with compact resolvent. This operator
enables the definition of distance between states. For the case of
the algebra of functions on a spin manifold, choosing for $D$ the
usual Dirac operator reproduces the usual metric distances, once
the equivalence between pure states and points is established.

Another important role played by the Dirac operator $D$ is in the construction
of the algebra of differential forms in the context of noncommutative geometry.
The key idea is to also represent differential forms as operators on ${\cal
H}$, on a par with $D$ and $\mathcal{A}$. We first define the (abstract) {\em
universal differential algebra of forms} as the $\zed$-graded algebra
\beq
\Omega^*\mathcal{A} = \bigoplus_{p\geq0}\Omega^p\mathcal{A}
\eeq
which is generated as follows:
\beq
\Omega^0\mathcal{A}=\mathcal{A}
\eeq
and $\Omega^1\mathcal{A}$ is generated by a set of abstract symbols $da$ which
satisfy:
\bea
d(ab) &=& (d a)b + a d b~,~~~ \forall ~a,b \in \mathcal{A}\ ~~~\mbox{(Leibnitz
Rule)}  \label{leib} \\
d(\alpha a + \beta b) &=& \alpha d a + \beta d b~,~~~ \forall ~a,b \in
\mathcal{A}~, ~~\alpha , \beta \in \complex\ ~~~\mbox{(Linearity)}  \nonumber
\eea
Elements of $\Omega^p\mathcal{A}$ are linear combinations of elements of the form
\beq
\omega=a_0da_1\cdots da_p \label{pforms}\eeq Because of
(\ref{leib}) a generic $p$-form can be written as a linear
combination of forms of the kind (\ref{pforms}), with $a_0$
possibly a multiple of $\mathbb{I}$ in the unital case. This makes
$\Omega^p\mathcal{A}$ a $\zed_2$-graded $\mathcal{A}$-module. The
graded exterior derivative operator is the nilpotent linear map
$d:\Omega^p\mathcal{A}\to\Omega^{p+1}\mathcal{A}$ defined by
\beq
d(a_0da_1\cdots da_p)=da_0da_1\cdots da_p
\eeq

We define a linear representation $\pi_D:\Omega^*\mathcal{A}\to{\cal B}({\cal H})$ of
the universal algebra of abstract forms by
\beq
\pi_D(a_0da_1\cdots da_p )= a_0[D,a_1]\cdots [D,a_p]
\eeq
Notice that $\pi_D(\omega)=0$ does not necessarily imply
$\pi_D(d\omega)=0$. Forms $\omega$ for which this happens are called {\em junk
forms}. They generate a $\zed$-graded ideal in $\Omega^*\mathcal{A}$ and have to be
quotiented out \cite{connes,landi}. Then the noncommutative differential algebra
is represented by the quotient space
\beq
\Omega_D^*\mathcal{A}=\pi_D\left[\Omega^*\mathcal{A}/(\ker\pi_D\oplus d\ker\pi_D)\right]
\label{OmegaD}\eeq
which we note depends explicitly on the particular choice of Dirac operator $D$
on the Hilbert space $\cal H$.

The algebra $\Omega_D^*\mathcal{A}$ determines a DeRham complex whose cohomology
groups can be computed using conventional methods. With this machinery, it
is also possible to naturally define (formally at this level) a
vector bundle $E$ over $\mathcal{A}$ as a finitely-generated projective left
$\mathcal{A}$-module, and along with it the usual definitions of connection,
curvature etc. However, in what follows we shall for
the most part use only the trivial bundle over a unital $C^*$-algebra $\mathcal{A}$.
For this we define a gauge group ${\cal U}(\mathcal{A})$ as the group of unitary
elements of $\mathcal{A}$,
\beq
{\cal U}(\mathcal{A})=\left\{u\in\mathcal{A}~|~u^\dagger
u=uu^\dagger=\mathbb{I}\right\} \label{unitarygp}\eeq
The presence of
one-forms is then tantamount to the possibility of defining a
connection, which is a generic Hermitian one-form
$A=\sum_ia_i[D,b_i]$, and with it a covariant Dirac operator
$D_A=D+A$.  The curvature of a connection $A$ is defined to be
\beq
F=[D,A]+A^2 \label{curv}
\eeq

Other structures of geometry are similarly rendered in an
algebraic form. Integrals become trace, with the role of measure
played by the inverse of the Dirac operator. The integral of a
function, seen as element of the algebra represented as operator
is a regularized trace called the {\it Dixmier trace}. Consider a
generic bounded operator $L$ on $\cal H$ of discrete spectrum with
eigenvalues $\lambda_n$ ordered according to their modulus and
counted with the appropriate multiplicities. The Dixmier trace
$\tr_\omega L$ is
\beq
\tr_\omega L=\lim_{N\to\infty}\frac{1}{\log N}\sum_{n=1}^N \lambda_n
\eeq
For the algebra of continuous functions on a $p$-dimensional compact manifold
$M$, this definition then yields \cite{connes}
\beq
\int_M f(x)~d^px =\tr_\omega~ f |D|^{-p} \label{intdirac}
\eeq
The differential structure of a manifold can be established by a
set of properties involving the algebra, the Hilbert space and the
$D$ operator. These axioms~\cite{Connesaxiom} provide an
algebraic characterization of manifolds in a completely algebraic
way.

The algebra $\mathcal{A}$, its representation on the Hilbert space
$\mathcal{H}$ and the $D$ operator provide the metric information
of the space, and together are called \emph{Spectral Triple}. Two
other operators, $J$ and $\chi$, encode the information about the
\emph{real} and \emph{chiral} structures respectively. In the case
of $\mathcal{A}$ being commutative this is just a rewriting of the
usual geometry of the underlying space, but the constructions can
be considered independently of the space, and therefore provide a
description of the noncommutative geometry.

With the progressive completion of this dictionary the emphasis
for the study of a geometrical structure passes from the
\emph{points} to algebra, or as seen from a physicist point of
view, to \emph{fields}. This research is giving several important
results in pure mathematics. But we now turn our attention and
consider some of the applications of these ideas to physics.

\section{Noncommutative geometry as a tool for\\ physics}

As we have seen, the need from a noncommutative geometry has deep
physical roots (quantum mechanics), moreover also its mathematical
tools naturally resonates with physicists. Starting from the late
eighties, noncommutative geometry, has been present in theoretical
physics. The definition of what is noncommutative geometry in
physics is however debatable, as the name has been attached to
different theoretical frameworks. A possible unifying character is
the idea that a modification of the structure of spacetime may be
important for the construction of physical theories. This
definition however leaves out some important related aspects, such
as fuzzy spaces and some applications of quantum groups. In the
following we will present some of the application of physics
without any pretense of completeness, neither in the choice of
topics, nor in their description.

\subsection{Connes approach to the standard model}

The standard model~\cite{Weinberg} of elementary particles and
their strong and electroweak interactions is a theory with
$SU(3)\times SU(2) \times U(1)$ gauge fields coupled to matter
fields in a suitable representation. It is one of the most
successful theories of last century and it has several
experimental confirmations in accelerator experiments. In
conjunction with general relativity it is the best framework for
understanding the universe after the Big Bang. The only missing
piece, the Higgs particle, is likely o be found at the {\sl Large
Hadron Collider} in the near future.

Attempts to go \emph{beyond the standard model} have not been
equally successful. First there were proposals of unification
based on theories similar to the standard model, with different
gauge groups containing $SU(3)\times SU(2) \times U(1)$, such as
$SU(5)$ or $SO(10)$. Such \emph{Grand Unified Theories} run into
difficulties, such as the proliferation of gauge bosons and the
incorrect lifetime of the proton. Models based on
\emph{supersymmetric} unification are still popular, but they lack
the experimental evidence of supersymmetric partners\footnote{This
too can change with the data of the large hadron collider.}. Connes
approach to gauge theories, based on noncommutative geometry,
provides a framework to \emph{understand} the standard model as
the ``electrodynamics'' of a noncommutative space. In this case
however the space is only ``almost'' noncommutative, in the sense
that the noncommutative algebra describing space is  the algebra
of functions over ordinary spacetime. The algebra is
noncommutative, but the points are still present. The model
is therefore to be seen at best as an effective theory.

The initial datum is that the geometry is encoded in the algebra,
the Hilbert space and $D$ operator, plus some other ingredients
like the representation of the algebra, the charge conjugation
operator and chirality. It is also possible to define in a
geometrical way the action of the theory, using only the spectral
data. Gauge transformation are unitary elements of the algebra and
define a covariant $D$ operator
\be
D_A=D+A
\ee
where $A$ is a self adjoint one-form, element of the space define
in \eqn{OmegaD}. It is possible to define the action of a
``noncommutative gauge theory''. The fermionic action is defined
as
\beq
S_F=\tr_\omega~\psi^\dagger D_A\psi=\langle\psi|D_A|\psi\rangle
\eeq
where the last equality gives the usual inner product on the
Hilbert space of square-integrable spinors, while the bosonic
action is purely spectral~\cite{ChamseddineConnes}
\beq
S_B=\tr\chi\left(\frac{D_A^2}\Lambda\right)
\eeq
with $\chi$ a smooth cutoff whose precise form is not important.
It is usually taken a smoothened version of the step function,
$\chi(x)\simeq 1$ for $x\lesssim 1$ and $\chi(x)\simeq 0$ for
$x\gtrsim 1$, and $\Lambda$ is the renormalization scale.

The remarkable fact is that, when one calculates this model for
the simple $C^*$ algebra of functions of spacetime valued in
diagonal $2\times 2$ matrices the resulting action describes a
spontaneously broken gauge theory with an Higgs
mechanism. This suggests the investigation of
other $C^*$-algebras which can describe the standard model. The
programme then is not to predict a Lagrangian, which on the
contrary is taken as input but to find a noncommutative geometry
which describes the standard model. The construction and
calculations to find the action are quite involved, but
straightforward, making use of heath kernel techniques.

The algebra $\mathcal{A}$ of the model is the tensor product of
the commutative algebra of functions on spacetime $C(M)$ times a noncommutative finite dimensional algebra
$\mathcal{A}=C_0(M)\otimes\mathcal{A}_F$. The latter is
necessarily a matrix algebra, and to reproduce the standard model
one takes
\be
\mathcal{A}_F=\complex\otimes\mathbb{H}\otimes M_3(\complex)
\ee
where $\mathbb{H}$ are the quaternions and $M_3(\complex)$ are
$3\times 3$ matrices. The Hilbert space correspondingly has a
continuous part, spinors on space time, and a finite part
comprising the 96 degrees of freedom of leptons and bosons
including colour, flavour, and helicity. The representation of
$\mathcal{A}$ on $\mathcal{H}$ is highly nontrivial and makes use
of the symmetry given by the charge conjugation operator $J$,
which also ha a continuous and a finite part. The $D$ operator as
well is composed of a finite and a continuous part:
$D=\gamma^\mu\del_\mu\otimes \mathbb{I}\otimes\gamma_5\otimes
D_F$, where $D_F$ encodes all of the mass and
Cabibbo-Kobayashi-Maskawa mixing of fermions (neutrinos included).
These are taken to be an \emph{input} and no effort is made in the
model to predict them. The details of the model, expecially with
regard to the representations of the algebra, the action of the
$D$ and $J$ operators and the way the algebra emerges, are
extremely complicated, nevertheless the calculation of the action
and its renormalization group analysis is straightforward. The
model has some predictive power of the mass of the Higgs particle,
which however depends on the nuances of it. Previous models, which
were mathematically simpler, and some problems with a doubling of
the degrees of freedom \cite{doubling} and the neutrino masses. It
reproduces the standard model with gravity and predicts a mass of
the Higgs at $170~GeV$, a value recently excluded by {\sl
Tevatron} data.

The latest model \cite{ChamseddineConnesMarcolli} is certainly the
most powerful and coherent one, but it assumes an almost
commutative geometry all the way to very high scale, and as a
consequence that the renormalization group analysis can be done in
the usual way. I still find astonishing the fact that a model
comes up with an Higgs mass "in the right ball park" from purely
geometric considerations, and while it is certainly not "the
final" answer, it remains a fascinating  model.

\subsection{The noncommutative geometry of strings}

In string theory spacetime plays a different role from a generic field theory. In the latter the spacetime is the arena in which the field live. String theory is a two-dimensional conformal field theory, and spacetime coordinates are the \emph{fields} of this theory, functions of the two coordinates of the world sheet. Quantization of these fields determines, for example, their number, i.e.\ the dimension of spacetime.  This different point of role that the coordinates of spacetime can be seen as an indication that one should expect to use a different geometry to describe it. Nevertheless for a some time the geometry used in string theory was the ``usual'' one.  
As far as I am aware, the first appearance the expression ``noncommutative geometry'' in a physics paper has been Witten's paper on open string field
theory~\cite{WittenSFT} in 1896. String fields are seen as maps from a string configuration
in space into complex numbers, with an enormous gauge symmetry
(reparametrisation). After gauge fixing the role of differential operator $\dd$
is played by the BRSt charge $Q$.

The coordinates of spacetime are however the fields of a \emph{free} string theory. Interactions  between strings are described by the insertion of appropriate \emph{vertex operators} on the world sheet. The generalization of the Lie algebra provided by vertex operators is an extremely active field of investigation in mathematics, with connections with conformal field theory, the monster group and several infinite dimensional algebras, such as Ka\c-Moody algebras. General introductions can be found in~\cite{Kac,gebert}.

The vertex operators can be see as operators on the Hilbert space of string states. The vertex operators are usually unbounded and must be smeared. This leads to the study of a noncommutative geometry of strings based on the algebra generated by them. The corresponding string spectral triple was introduce in~\cite{FrohlichGawedzki, LizziSzaboPRL, LizziSzabo} and called the  \emph{Fr\"ohlich-Gaw\c edzki} spectral triple. The third ingredient is the Dirac operator. This is built from the fundamental fields (the coordinates of spacetime). In the case of toroidally compactified strings we have that these are the Fubini-Veneziano fields:
\be
X^\mu_\pm(\tau\pm\sigma)=x^\mu_\pm+g^{\mu\nu}p_\nu^\pm(\tau\pm\sigma)
+\sum_{k\neq0}\frac1{ik}~\alpha^{(\pm)\mu}_k~\e^{ik(\tau\pm\sigma)}
\ee
where the suffix $\pm$ refers to left and right movers, and with a conformal transformation we have mapped the world sheet, a cylinder of coordinates $\sigma=\sigma+2\pi$ and $\tau$ into a complex plane, with the transformation $z_\pm=\e^{(-\ii\tau\pm\sigma)}$. For tori with equal radii, we have that the the center of mass of the string is 
\be
x^\mu=x^\mu_++x^\mu_-
\ee
is the position of the center of mass of the string, while its momentum of the string and winding are
\bea
p_\mu=p_\mu^++p_\mu^-\nonumber\\
w_\mu=p_\mu^+-p_\mu^-
\eea
Note that for a toroidally compactified string the momentum is quantized, on a par with the winding number. The eigenvalues of momentum are proportional to the inverse of the radius, while those of he winding are directly proportional to the radius. The zero-modes
$x^\mu_\pm$ and $p_\mu^\pm$ are canonically conjugate
variables upon quantization.

We can define Dirac operator as
\bea
D&=&D_++D_-\label{Diracplus}\\
D_\pm&=&\gamma_\mu\partial_\pm
X_\pm^\mu(\tau\pm\sigma)
\eea
and extract ordinary spacetime as the low energy subspectral triple obtained eliminating all oscillator and winding modes. The latter elimination is a low energy limit for the case in which the compacification radius of the torus is very large, compared with the string scale, which in turn is thought to be of the order of Planck's length. In this case the spectrum of momentum is effectively continuous, since the eigenvalues are very close to each other, being proportional to a very small quantity. The only relevant vertex operators are the one corresponding to the ground state of the string, without oscillatory modes, they commute and, for the emission of a string of momentum $p$ can be expressed as
\be
V_p=\e^{\ii p_\mu x^\mu}
\ee
and they generate the algebra of functions on space time, since in this case smearing is equivalent to Fourier transform with a continuous function.

Things are different in the case of very small radius. Notice the plus sign in ~\eqn{Diracplus}, we could have used equally well the minus sign, in fact the spectral triple is invariant for the transformation 
\be
z_\pm\,\to \pm z_\pm
\ee
which can be implemented by a unitary transformation, and therefore is a gauge transformation. The transformation exchanges position with winding, $D$ with $D_w=D_+-D_-$ and $x$ with $x_w^\mu=x^\mu_+-x^\mu_-$. Since $x_w^\mu$ is canonically conjugated with $w_\mu$, which is with a nearly continuous spectrum for very small radius. In this limit the commutative subtriple is generated by the vertices
\be
V_w=\e^{\ii w_\mu x_w^\mu}
\ee
and the resulting spacetime is the same that we should have obtained with the transformation $R\to\frac{\ell_p^2}{2\pi R}$.
This is the celebrated T-duality of closed strings (for a review see~\cite{tdualrev}). We have seen that in the contest of the noncommutative geometry of strings it appears  as a gauge transformation.

In the case for which the compactification radius is of order one (in the string scale), then , still neglecting fluctuations, both momentum and winding will play a role, and the ``low'' energy algebra generated by the vertex operator corresponding to the low lying states do not any longer commute among themselves~\cite{LandiLizziSzabo}. What happens is that
\be
V_pV_w=\e^{\ii\omega^{\mu\nu}p_\mu w_\nu}V_wV_p
\ee
where $\omega$ is an antisymmetric matrix. This is a structure known as the \emph{noncommutative torus}, and is nothing but the exponentiated version of the canonical commutation relation.

Canonical commutation relations for string coordinates appear also in the case of open strings~\cite{Schomerus, SeibergWittenNCG}, in a particular limit in the presence of branes with a background field.  What happens in this case is that effectively the coordinates of spacetime do not commute giving a commutation relation similar to that of quantum mechanics
\be
[x^\mu,x^\nu]=\ii\theta^{\mu\nu}
\ee
Seiberg and Witten~\cite{SeibergWittenNCG} proposed to describe the field low energy corresponding to this string configuration encoding the noncommutativity of space into a deformed product. They thus showed that, at least in a particular limit, string theory are described by a field theory on a noncommutative space. This I discuss in the next section.

\subsection{Field theory on a noncommutative space}

The possibility to use canonical commutation relations among coordinates to solve the localization problem described in section~\eqn{se:need} was first introduced by Doplicher, Fredenhagen and Roberts~\cite{DoplicherFredenhagenRoberts}, but without any doubt it received a great impulse after the work of Seiberg and Witten~\cite{SeibergWittenNCG}. They also indicated that a way to describe a field theory on a noncommutative space was to use a deformation of the product. This had been introduced by Weyl~\cite{Weyl}, Gr\"onewold~\cite{Gronewold} and Moyal~\cite{Moyal} as a method of quantization. the idea is that a quantization of a space (be it phase or configuration space) is achieved introducing a new associative, but noncommutative product, depending on a small quantity. The product is required to reproduce the nontrivial commutation relations among the generators and the ordinary, commutative product in the limit of vanishing parameter. The small parameter is obviously $\hbar$ for the quantization of phase space, while for space time it is an antisymmetric matrix which is usually called $\theta^{\mu\nu}$.

We define therefore the Moyal product between two functions as
\be
f\star g=f e^{\frac
i2\theta^{\mu\nu}\overleftarrow{\del_\mu}\overrightarrow{\del_\nu}}g \label{Moyalasym}
\ee
where the symbol $\overleftarrow{\del_\mu}$ indicates that the derivative is acting on the left. Applying the product to the coordinate functions we obtain the canonical commutation relation
\be
x^\mu\star x^\nu - x^\nu\star x^\mu=\ii\theta^{\mu\nu} \label{commx}
\ee
The definition~\eqn{Moyalasym} is not the most robust for the product, but it is well defined on polynomials and Schwarzian functions, mapping them again into polynomials and Schwarzian functions respectively. It may be considered an asymptotic expansion~\cite{VarillyGraciaBondiaasymexp} of more solid integral expressions. In particular its Fourier transform is a twisted convolution, in $d$ dimensions we have
\be
\widetilde{(f\star g)}(p)=\frac{1}{(2\pi)^{\frac d2}} \int\dd^d q \tilde f(q)\tilde g(p-q) \e{p_\mu\theta^{\mu\nu}q_\nu} \label{twistedconvo}
\ee
The antisymmetric matrix $\theta^{\mu\nu}$ is considered to be composed of small dimensionful parameters, expanding the exponential in~\eqn{Moyalasym} we see that
\be
f\star g= fg + \{f,g\} + O(\theta^2)
\ee
where $\{f,g\}=\theta^{\mu\nu}\del_\mu f \del_\nu g$ is the Poisson structure defined by the antisymmetric matrix $\theta^{\mu\nu}$. To order zero the product is the ordinary commutative one, and the $\star$ product is a \emph{deformation} of it. The Moyal product, seen in phase space, is actually the archetype of a deformed product providing a quantization of a classical Poisson structure~\cite{Bayenetal1,Bayenetal2}. It is possible to prove~\cite{Kontsevich} that, at least at the level of formal series in the coordinate functions, given a Poisson structure it is always possible to find a $\star$ product which reproduces the Poisson structure to first order. The procedure to prove this is not simple, mainting associativity to all order is extremely difficult, but the result is very important in quantum mechanics. It means that all classical systems defined by a  Poisson bracket, can be (at least formally) quantized.

For the quantization of spacetime the presence of a $\star$ products means that we have encoded the noncommutativity in the algebra of the fields.  We have therefore a procedure to define field theory on noncommutative spaces\footnote{For a review of field theory on noncommutative spacs see~\cite{Szabo}.}. This is accomplished rewriting the action of a field theory substituting the usual product with the deformed product. For example consider a scalar field theory with action
\be
S=\int d^dx \del_\mu\varphi\star\del^\mu\varphi+ m^2
\varphi\star\varphi
+\frac{g^2}{4!}\varphi\star\varphi\star\varphi\star\varphi \label{phiaaction}
\ee
Note that the Moyal product has the property that 
\be
\int d^dx f\star g =\int d^dx f g  
\ee
therefore the $\star$ in the first two terms of the action~\eqn{phiaaction} are actually redundant. As a consequence the free theory is the same in the commutative and noncommutative cases. The vertex~\cite{Filk} is not the same however, and it acquires a phase:
\be
V=(2\pi)^4g\delta^4\left(\sum_{a=1}^4
{k_a}\right)\prod_{a<b}\e^{-\frac i
2\theta^{\mu\nu}{k_a}_\mu{k_b}_\nu}
\ee
where the $k_a$ are the incoming momenta. The presence of the phase, which is a consequence of the phase in~\eqn{twistedconvo}, alters the properties of the loop diagrams. The first consequence is that the vertex is not anymore invriant for exchange of the momenta, it is unchanged only for cyclic permutations, therefore for example the one-loop corrections to the propagator in Fig.~{planardiagram}
\begin{figure}[htbp]
\epsfxsize=4.5 in
\bigskip
\centerline{\epsffile{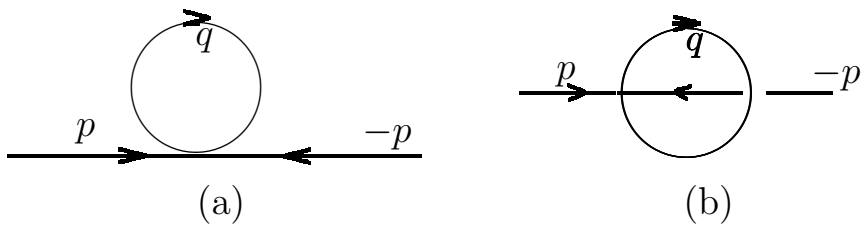}}
\caption{\sl The planar (a) and nonplanar (b) one loop correction
to the propagator} \label{planardiagram}
\end{figure}
are different in the two cases. The Green's functions are (with the appropriate combinatorial factors)
\bea
G^{(2)}_{\mathrm P} &=&  - \ii \frac{g}{3}
\int\frac{\dd^dq}{(2\pi)^3}\frac{1}{(p^2-m^2)^2(q^2-m^2)}\nonumber\\
G^{(2)}_{\mathrm NP} &=&  - \ii \frac{g}{6}
\int\frac{\dd^dq}{(2\pi)^3}\frac{\e^{\ii p_\mu\theta^{\mu\nu}q_\nu}}{(p^2-m^2)^2(q^2-m^2)}
\eea
While for the planar case (a) the contribution is the same as in the the commutative case, the phase of in the nonplanar case (b) eliminates the \emph{ultraviolet} divergences, at the price of an \emph{infrared} divergent behaviour. This is the ultraviolet/infrared mixing~\cite{MvRS}. Heuristically this can be seen as a consequence of the uncertainty in spacetime caused by the nonzero commutator~\eqn{commx}. The short distance behaviour in the direction $x^\mu$ has consequences in the long distances of $\theta_{\mu\nu}x^\nu$.

Other theories can also be studied on noncommutative spaces, notably abelian and nonabelian gauge theories, which for lack of space we will not discuss presently and refere again to the literature, for example the references and citations of~\cite{Szabo}.

\section{Deformed symmetries of noncommutative field theory}

In this section we will see how the deformation of spaces induced by noncommutative geometry requires a deformation of the symmetries as well. We will then concentrated on a particularly relevant kind of deformation, that induced by a twist.

\subsection{Deformed symmetries and Hopf algebras}

The presence of the quantity $\theta^{\mu\nu}$ necessarily breaks Lorentz invariance, selecting two directions in space given by the two vectors $\theta^{0i}$ and $\varepsilon_{ijk}\theta^{jk}$. In the case of a constant $\theta$ translations remain a symmetry. The presence of a ``fundamental'' breaking of Lorentz symmetry can be seen with interest, and we will in fact discuss the possibility that precisely the breaking if the symmetry may help in he measurement of $\theta$. On the other side there is also the point of view that a fundamental theory, which has among its main motivation quantum gravity, should maintain the main relativistic symmetry.

One possibility is that the theory is still symmetric, but in a ``deformed'' way, namely invariance under an Hopf rather than a Lie algebra. Let us give a few pointers on these deformed symmetries, again we refer to the literature for details. A standard reference is~\cite{Majidbook}\footnote{The theory is usually also referred as ``quantum groups''.
This meaning of quantum has nothing to with quantization, as we use in the rest of this review, and we avoid using it, preferring the nomenclature Hopf algebras}.

Consider functions on a manifold $M$ which happens to be the space of parameters of a Lie group. The $C^*$-algebra $C(M)$ of functions on this manifold is commutative (even if the group is not) and sufficient to reconstruct the space. A group manifold has of more structure. A particular point is the identity $e$, given two points $g$ and $g'$ we can find a third, the product of the two $gg'$, there is the inverse of every point with the usual properties, the product is associative. it is possible to encode this structure on the space of functions via the three maps, \emph{coproduct} $\Delta:\,C(M)\otimes C(M)\to C(m)$, \emph{counit} $\epsilon:\,C(M)\to\complex$  and \emph{antipode} $S:\,C(M)\to C(M)$. Given a function $a(g)$ we defines 
\bea
\Delta(a)(g,g')&=&a(gg')\nonumber\\
\epsilon(a)&=&a(e)\nonumber\\
S(a)(g)&=&a(g^{-1})
\eea

Consider then the Lie algebra $g$ of the group $G$ and the pairing obtained acting with the element of the Lie algebra $L\in g$ seen as vector field, acting on the function and evaluating it on the identity of the group.
\be
\langle L,f\rangle=L(f)\bigg|_e
\ee
We can extend this to the universal enveloping algebra of the Lie algebra $U(g)$, that is the algebra of polynomials in the elements of $g$, with unity $I$, and modulo the commutation relations. We can give an Hopf algebra structure to $U(g)$, defining coproduct, counit and antipode as follows:
\bea
\Delta(L)=L\otimes I+I\otimes L &~~& \Delta(I)=I\otimes I\nonumber\\
\epsilon(L)=0 &~~& \epsilon(I)=1\nonumber\\
S(L)=0 &~~& S(I)=I
\eea
and then extending it by the requirement $\Delta(LL')=\Delta(L)\Delta(L')$. The compatibility is given by the following:
\bea
\langle L ,ff'\rangle&=&\langle \Delta(L),f\otimes f'\rangle\nonumber\\
\langle L L',f\rangle&=&\langle L\otimes L',\Delta(f)
\eea
note that the first of these relations is the Leibnitz rule.

The algebra of functions on the group is commutative, and this is reflected in the fact that the Hopf algebra of $U(g)$ is \emph{cocommutative}, i.e.\ $\Delta(L)(f\otimes f')=\Delta(L)(f'\otimes f)$. If we now consider a deformation of the functions on the group we must correspondingly deform the Hopf algebra of $U(g)$. We could say we have a quantum group, had not we decided not to use this terminology.

For the case  at hand consider the algebra of functions on spacetime and the action of the universal enveloping algebra of the Poincar\'e Lie algebra. Note however that the Leibnitz rule is no longer valid with a $\star$ product:
\be
M_{\mu\nu} (f\star f')\neq f M_{\mu\nu} f' + (M_{\mu\nu}f)f'
\ee
where $M_{\mu\nu}=x_\mu\del_\nu-x_\nu\del_\mu$. The noncommutative product induces a deformation of the Hopf algebra of the Poincar\'e generators. We will write down the deformed coproduct in a moment, because it will be easier to calculate and understand it after the introduction of another structure: the \emph{Drinfeld twist}.

\subsection{Twisted symmetries}
In this section we present a simple yet profound way to see the $\star$ product, it is based on the the \emph{Drinfeld twist}~\cite{Drinfeld}. In this context was introduced in~\cite{Oeckl, Helsinki, Wess, aletWess}. Details of the construction can be found in the forthcoming book~\cite{WSS}. For this particular case we define the twist as the following map from two copies of the algebra of functions on spacetime into itself as follows
\be
\mathcal{F}=\e^{-\frac\ii2\theta^{\mu\nu}\del_\mu\del_\nu}: C(M)\otimes C(M) \to C(M)\otimes C(M)
\ee
the twist can be seen as a deformation of the tensor product. Consider the usual commutative product as a map $C(M)\otimes C(M) \to C(M)$:
\be
m_0(f\otimes g)= fg
\ee
then we can define the $\star$ product as the composition of $m_0$ and the inverse of the twist\footnote{The reason the inverse of $\mathcal{F}$ appears in the relation for the product is historical, it has no particular meaning.}
\be
m_\star(f\otimes g)=m_0\circ\mathcal{F}^{-1}((f\otimes g)=f\star g
\ee
The twist defines as well a deformation of the coproduct structure of the Poincar\'e Hopf algebra and one defines:
\be
\Delta_{\mathcal{F}}=\mathcal{F}\Delta\mathcal{F}^{-1}
\ee
where $\Delta_0$ is the undeformed coproduct. The new deformed Hopf algebra therefore has the same Lie algebra structure, the same counit and antipode, but a different coproduct for the rotations and boosts part (since translations commute with the twist, their coproduct is unchanged). The new coproduct is
\bea
\Delta_{\mathcal{F}}(M_{\mu\nu})&=&M_{\mu\nu}\otimes I +I\otimes M_{\mu\nu}\\ && -\frac\ii2 \theta^{\alpha\beta}  \left(\left(\eta_{\alpha\mu}P_\nu - \eta_{\alpha\nu}P_\mu\right)  \otimes P_\beta  +P_\alpha\otimes \left(\eta_{\beta\mu}P_\nu - \eta_{\beta\nu}P_\mu\right)\right)\nonumber
\eea
While the Hopf algebra is different, the Lie algebra part of it remains unchanged, this means for example that the Casimir invariants are the same, which in turn implies that the Wigner classification of particles still holds.

One can then take the point of view~\cite{Aschieri, AschieriLizziVitale} that what is fundamental is the deformation of the tensor product, and that the twist induces a deformation of all products present in the theory, i.e.\ given a bilinear map $\mu: X\otimes Y\to Z$, where $X,Y$ and $Z$ are spaces on which a representation of the translations, and hence of the twist $\mathcal F$, we can define a twisted product
\be
\mu_\star= \mu\circ\mathcal{F}^{-1}
\ee
in particular when the three states are all the space of functions and $\mu=m_0$ we obtain again the Moyal product. We can apply this procedure consistently from classical mechanics to field theory to the calculation of the S-matrix~\cite{GalluccioLizziVitale}.

The physical significance of the fact that we still have a theory which is Poincar\'e symmetric, albeit under a deformation of the Hopf algebra, is not totally clear, and has lead to controversies. Two extreme positions are possible: a twist deformation in unobservable~\cite{Fiore}, or on the contrary the breaking of Lorentz invariance is unavoidable in these theories~\cite{ChaichianSalminenTureanu}. What is probably lacking at this time is a fully developed measurement theory in the presence of these deformations.

\section{Noncommutative geometry and the real\\ world}

If we take seriously the fact that the world is described by some 
kind of noncommutative field theory which are the consequences? And in particular, given the description of the theory by the Moyal product, how do we measure $\theta^{\mu\nu}$, a quantity of the order
of $\ell_P^2$?

One first possibility is to consider the consequences in the scattering of elementary particles. The presence of the $\star$ product has the consequence of altering the vertex and therefore the Green's functions of the theory. Another consequence is the fact that, due to the noncommutative product among the fields, also a U(1) theory is ``non abelian''. This gives new interactions in gauge theories. In fact in noncommutative gauge theories the curvature tensor is
\be
F_{\mu\nu}=\del_\mu A_\nu-\del_\nu A_\mu- e (A_\mu\star A_\nu - A_\nu \star A_\mu)
\ee
and the last term is nonvanishing also for a scalar potential. The problem is that if $\theta^{\mu\nu}$ has a fixed direction in space, then the experimental effects are washed out by the rotation of the earth. Therefore it is best to look for decays which would be forbidden, for example the decay $Z\to\gamma\gamma$~\cite{Trampetic1}. There has been some work on noncommutative phenomenology, starting from the Opal collaboration~\cite{Opal}. More recent work has been done for example by in~\cite{OhlReuter,AlboteanuOhlRuckl,Trampetic2}. A recent review is~\cite{Alboteanu}.

Another source of phenomenologically viable observations of noncommutativity is the early universe. Even a Planck size noncommutativity would be relevant in the early universe. At this level, and as suggested by string theory,
$\theta^{\mu\nu}$ is a background quantity. Since it is an antisymmetric tensor it selects
two directions in space (analog of electric and magnetic fields from $F_{\mu\nu}$).
Their presence breaks Lorentz invariance and the noncommutativity
will have left its imprinting in the early universe, and its
consequences are thereafter frozen by inflation~\cite{ChuGreeneShiu}. These consequences could be observed with present, or more likely future, surveys of the cosmic microwave background. The problem is that the Moyal product is made for flat
coordinates. The construction of {associative} deformed products
is not simple (Kontsevich has won a field medal building them!).
One cannot simply substitute, say, the partial derivatives in the
definition with covariant derivatives, in this case the product would be nonassociative, and this would render the definition of a field theory very problematic.

In~\cite{LMMP} we considered
the field theory of a field which causes inflation to be deformed
by a star product. We define a curved star product to first order in
$\theta=\theta^{12}=1/\Lambda a^2$, where $a$ is the
usual scale factor of the universe and $\Lambda^{-1}$a
noncommutativity scale. Notice that the fact that $\theta$
is a tensor chooses a direction (12 in our case). With this choice and a Moyal product defined for comoving coordinates with covariant
derivatives, nonassociativity is fourth order effect, and one can
study the corrections to the inflaton action to low order in $\theta$.  
The corrections are such that the background evolves in the usual
way, but the fluctuations change. As a consequence the quadrupole in the cosmic microwave background will show an effect.

Just for completeness let me
show the extra terms (more details in~\cite{LMMP}):
\begin{equation}
\delta S_V = -\frac{m^2}{16} \, \int d^4 x \, a^3
\frac{1}{\Lambda^4} \left( \frac{\dot{a}}{a} \right)^2 \left(
\partial_1 \phi \partial^1 \phi + \partial_2 \phi \partial^2 \phi
\right), \label{delv}
\end{equation}
and
\begin{eqnarray}
\delta S_K &=& \frac{1}{32} \int d^4 x \sqrt{-g} \, \Theta^{\mu
\nu} \Theta^{\rho \sigma} \left( D_\rho D_\tau \phi \right) \left(
\left[ D_\mu , D_\nu \right] D_\sigma D^\tau \phi \right)
\nonumber\\
&=& \frac{1}{16} \int d^4 x \, a^3 \frac{1}{\Lambda^4} \left(
\frac{\dot{a}}{a} \right)^2 \biggl[ \partial_m \partial_0 \phi
\partial^m \partial^0 \phi + \partial_m \partial_i \phi \partial^m
\partial^i \phi -
\nonumber\\ &&
           \hphantom{\frac{1}{16} \int d^4 x \, a^3 \frac{1}{\Lambda^4} \left(
                     \frac{\dot{a}}{a} \right)^2 \biggl[}\!
- 2 \frac{\dot{a}}{a} \partial^m \phi \partial_0 \partial_m \phi +
\left( \frac{\dot{a}}{a} \right)^2 \partial_m \phi \partial^m \phi
\biggr] \, , \label{delk}\nonumber\\
\end{eqnarray}
where $\phi$ is the inflaton field and $m=1,2$. The consequence is a signature in the cosmic microwave background which shows up as a quadrupole correlation, which are encoded in the power spectrum with an orthogonality relation between the $a_{lm}$, the spherical harmonics coefficients of the power spectrum. In the usual case it results
\be
\vev{a^*_{lm}a_{l'm'}}\propto\delta_{ll'}\delta{mm'}
\ee
In the presence of the noncommutative product terms there appear also terms correlating $a_{lm}$ with $a_{l\pm 2 \, m}$

There is more work of a cosmological nature investigating the consequences of noncommutativity of various kind. A review is~\cite{Brandenberger}. The problem is that however it is not easy to distinguish the predictions coming from the presence of the Moyal product from these coming from a generic breaking of the Lorentz symmetry. Moreover these predictions are very much dependent on the specific form of the product, which may be just a first approximation of a more general product with $\theta$ nonconstant .

As we mentioned the construction of a noncommutative, associative product is not an easy task, making it `realistic'' is even more difficult. There are more noncommutative spaces that have been used to describe spacetime, and we cannot discuss all of them for lack of space. It is important however to at least mention $\kappa$-Minkowski spacetime~\cite{LRNT,LNR}. This space is characterized by the following commutation rule\footnote{The quantity $\lambda=\frac1\kappa$ is the small parameter of this theory, and it plays the same role $\theta$ plays for the Moyal product.}:
\bea
[x_i,x_0]&=&i\lambda x_i,\nonumber\\ {}[x_i,x_j]&=&0
\eea
and it emerges as the space dual to another Hopf algebra, called $\kappa$-Poincar\'e. Deformed products, analogs of the $\star$ prodcut can be built considering the system as a deformation of a Weyl system~\cite{Vivaldi}. A review of this space with an emphasis on symmetries is~\cite{Agostini}.

The commutation relations for $\kappa$-Poincar\'e are:
\bea
{}[P_\mu,P_\nu]&=&0\nonumber\\ {}[M_i,P_j]&=&i\epsilon_{ijk}P_k
\nonumber\\{}[M_i,P_0]&=&0\nonumber\\
{}[N_i,P_j]&=&-i\delta_{ij}\left(\frac{1}{2\lambda}(1-e^{2\lambda
P_0})+\frac{\lambda}{2} P^2\right)+i\lambda P_iP_j\nonumber\\
{}[ N_i,P_0] &=&iP_i\nonumber\\
{}[M_i,M_j]&=&i\epsilon_{ijk}M_k\nonumber\\
{}[ M_i,N_j] &=&i\epsilon_{ijk}N_k\nonumber\\
{}[N_i,N_j] &=&-i\epsilon_{ijk}M_k
\eea
All these commutation relations become the standard ones for
$\lambda\to 0$. The bicrossproduct basis is peculiar as
$\kappa$-Poincar\'e acts \emph{covariantly} on a space that is
necessarily deformed and noncommutative. This is a consequence of
the non cocommutativity of the coproduct which, always in the
bicrossproduct basis, reads:
\bea
\Delta P_0&=& P_0\otimes 1 + 1 \otimes P_0\nonumber\\
 \Delta M_i&=& M_i\otimes 1 +1 \otimes M_i\nonumber\\
 \Delta P_i&=&P_i\otimes 1+e^{\lambda P_0}\otimes P_i\nonumber\\
  \Delta N_i
&=&  N_i\otimes 1+e^{+\lambda P_0}\otimes
N_i+\lambda\varepsilon_{ijk}P_j\otimes M_k \label{coproducts}
\eea
The Casimir of this quantum group provide a deformation of the
Energy-Momentum dispersion relation and this could be used to
explain for example $\gamma$-ray bursts~\cite{Amelinogamma}. The
problem is that, being the commutation relations nonlinear,
nonlinear changes of coordinates are allowed, and therefore these
dispersion relations become basis-dependent, and the discussion has to move on the ``natural'' choice for a basis.

\section{Conclusions}

In this talk I have attempted a description of a variety of efforts to understand the structure of spacetime using a ``bag of tools'' called noncommutative geometry. The question is: are we using the correct bag of tool, and withis that bag, are we using the right tool?  And how do we know?

My personal conclusion is that at the Planck scale the geometry of spacetime is not the usual one, and that the mathematical structure describing it must be some form on nocommutative geometry. I am also convinced that deformed symmetries will play a central role. I am less convinced that we have already got the right kind of noncommutative geometry. In other words, I think we have the right bag of tolls, but we have not yet the right tool in our hands. In particular we still need a better physical understanding of physics to proceed to a complete description. This description will probably part of a larger framework, like string theory or loop quantum gravity, or a novel idea.
Fortunately we can expect some input from experiments and
observations: large hadron vollider, cosmic microwave background precision measurement, ultra high energy cosmic rays. Our best hope is that nature itself will give us some other clue.

\subsubsection*{Acknowledgments}
It is a great pleasure  to thank Joseph Kouneiher, Fr\'ed\'eric H\'elein, Volodya Roubtsov and C\'ecile 
Barbachoux for organizing such a stimulating workshop.

\end{document}